\documentclass[twocolumn,showpacs,preprintnumbers,amsmath,amssymb,prl,superscriptaddress]{revtex4-1}

\usepackage{graphicx}
\usepackage{dcolumn}
\usepackage{bm}
\usepackage{epsfig}
\usepackage{natbib}
\usepackage{color}
\usepackage{ulem}
\usepackage{enumerate}

\newcommand{\Bperp}{B_{\bot}} 
\newcommand{\Bparallel}{B_{\|}} 
\newcommand{\Btotal}{B_\text{total}} 
\newcommand{\me}{m_\text{0}} 
\newcommand{\lB}{l_\text{B}} 
\newcommand{\Rxx}{R_\text{xx}}  
\newcommand{\Rxy}{R_\text{xy}}  

\begin{document}

\title{Polarization dependent Landau level crossing in a two-dimensional electron system in MgZnO/ZnO-heterostructure}
\author{D.~Maryenko}
\email{maryenko@riken.jp}
\affiliation{RIKEN Center for Emergent Matter Science(CEMS), Wako 351-0198, Japan}
\author{J.~Falson}
\affiliation{Department of Applied Physics and Quantum-Phase Electronics Center (QPEC), The University of Tokyo, Tokyo 113-8656, Japan}
\author{Y.~Kozuka}
\affiliation{Department of Applied Physics and Quantum-Phase Electronics Center (QPEC), The University of Tokyo, Tokyo 113-8656, Japan}
\author{A.~Tsukazaki}
\affiliation{Institute for Materials Research, Tohoku University, Sendai 980-8577, Japan}
\author{M.~Kawasaki}
\affiliation{RIKEN Center for Emergent Matter Science(CEMS), Wako 351-0198, Japan}
\affiliation{Department of Applied Physics and Quantum-Phase Electronics Center (QPEC), The University of Tokyo, Tokyo 113-8656, Japan}

\today
\begin{abstract}
We report electrical transport measurements in a tilted magnetic field on a high-mobility two-dimensional electron system confined at the MgZnO/ZnO heterointerface. The observation of multiple crossing events of spin-resolved Landau levels (LLs) enables the mapping of the sequence of electronic states. We further measure the renormalization of electron spin susceptibility at zero field and the susceptibility dependence on the electron spin polarization. The latter manifests the deviation from the Pauli spin susceptibility. As the result, the crossing of spin-resolved LLs shifts to smaller tilt angles and the first Landau level coincidence event is absent even when the magnetic field has only a perpendicular component to the 2DES plane.
\end{abstract}


\maketitle
A magnetic field applied to a two-dimensional electron system (2DES) opens several energy gaps in the electronic spectrum. In a single particle picture, two fundamental energy scales dominate in the energy spectrum. One is the Zeeman gap ($E_\text{z}$), corresponding to the lifting of the spin degeneracy and given by $g^{\ast}\mu_B \Btotal$, where $g^{\ast}$ is the Land$\acute{e}$ $g$-factor, $\mu_B$ is the Bohr magneton and $\Btotal$ is the total magnetic field applied to the system. The other is the cyclotron gap ($E_\text{cyc}$), which reflects the quantization of electron cyclotron motion  by the formation of Landau levels (LLs) with energy $\hbar\omega_c(N+1/2)$, where $N$ is the Landau level index and $\omega_c$=$e\Bperp/(m^{\ast}\cdot \me)$ is the cyclotron frequency given by the magnetic field component normal to the 2DES plane, that is $\Bperp$=$\Btotal\cos(\theta)$. Here, $\theta$ is the angle between the direction of $\Btotal$ and normal to the sample, $\hbar$ is the reduced Planck constant, $e$ is the elementary charge and $m^{\ast}$ is the effective mass of the charge carrier in units of the electron free mass $\me$. The ratio of number of electrons to the degeneracy of the discrete energy levels, given by the number of magnetic flux threading the sample $e\Bperp/h$, defines a filling factor $\nu$ of this electron energy ladder and indicates how many of these levels are occupied. 

The two energy scales can be tuned relative to each other by tilting the sample in the magnetic field and the electron states at different Landau levels and with opposite spin orientation can eventually be brought to the coincidence. This method has been applied for various 2DESs to evaluate their electron spin susceptibility $\chi$, a fundamental property in condensed matter physics that describes the response of spin polarization $P$ to changes of the applied magnetic field $\Btotal$ ~\cite{FirstTiltFieldExperiment, Nicholas1988, Zhu2003, Vakili2004, SiInversionRotation, Shashkin2001, Shashkin2003, Tutuc2002,Vitkalov2001,InAsRotation}. It relies, however, on a linear dependence of $P$ on $\Btotal$, i.e. Pauli spin susceptibility:
\begin{equation}
P=\frac{\chi}{n}\Btotal=\frac{e}{2h}\frac{g^{\ast}m^{\ast}}{n}\Btotal
\end{equation}
with $n$ being the charge carrier density. However, such a linearity is not a given in a system with strongly interacting electrons. Rather $\chi$ depends on a net spin polarization $P$ and therefore one can question the reliability of coincidence method for evaluating the spin susceptibility \cite{DasSarma2006}. A high electron mobility MgZnO/ZnO heterostructure is a particular suitable system to probe this concept~\cite{IQHE_ZnO,FQHE_ZnO}. Firstly, the cyclotron and Zeeman gaps are comparable in size to each other, i.e. $E_\text{cyc}/E_\text{z}\approx3.3$ at $\theta=0$, when considering the band electron mass value ($m^{\ast}$~=~0.3) and the band electron Land$\acute{e}$ $g$-factor ($g^\ast$~=~2). Therefore, level coincidences may be achieved at moderate tilt angles and consequently, multiple coincidence positions, signifying a large range of polarizations, may be observed. This too may allow the mapping of these transitions and the polarization. 
Such a mapping will provide a more detailed information about the index dependence of each Landau level crossing. Secondly, the 2DES in MgZnO/ZnO heterointerface is a strongly interacting system with a reported enhancement of both electron mass and Land$\acute{e}$ $g$-factor~\cite{TsukazakiPRB, KozukaPRB, MaryenkoPRL}. Thus, this system ought to show a pronounced spin-susceptibility dependence on the electron polarization.

\begin{figure*}[!th]
\includegraphics{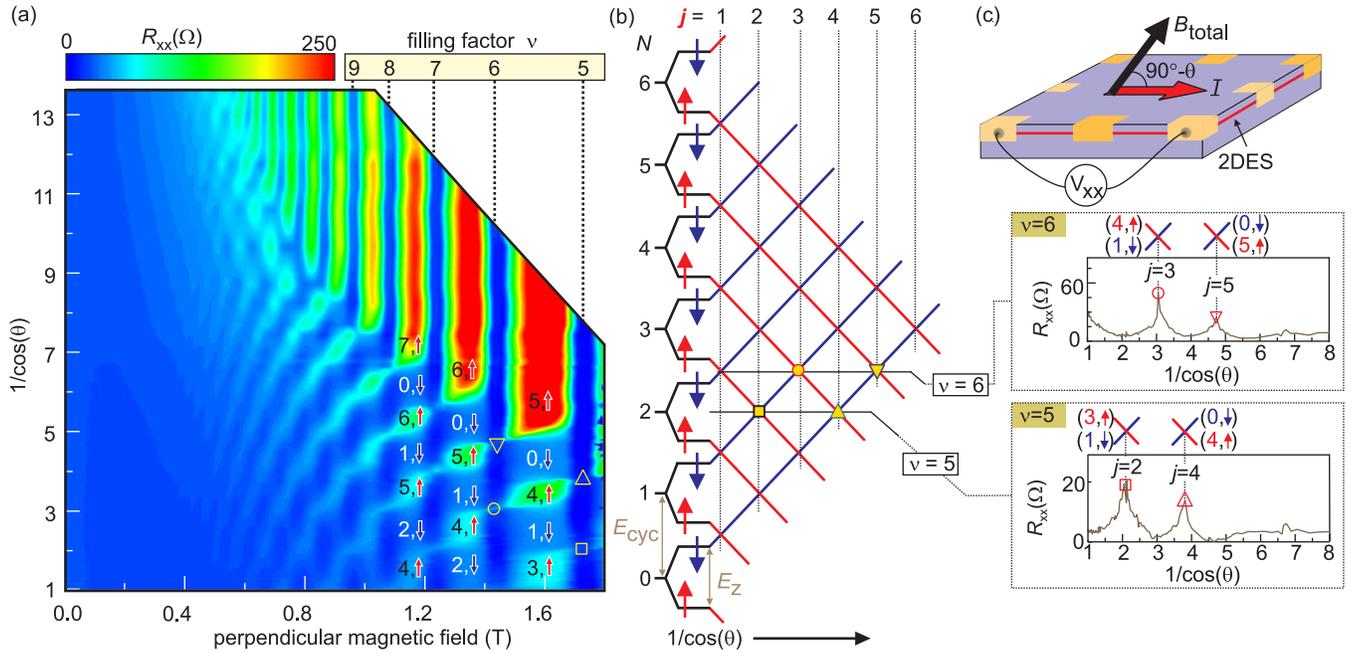}
\caption{\label{Fig1} (a) Color rendition plot of the longitudinal resistance $\Rxx$ as a function of the perpendicular magnetic field and $1/\cos(\theta)$ with $\theta$ being the tilt angle. Indicated are the electron states ($N,\uparrow/\downarrow$) at the chemical potential. (b) Scheme of the spin-resolved Landau level crossings in a magnetic field assuming a fixed component of the magnetic field perpendicular to 2DES plane (left panel). $j$ indicates the index of the level crossing event. $\Rxx$ traces are given for $\nu=5$ and 6 as a function of $1/\cos(\theta)$. (c) Schematic of the sample. The sample is tilted in the magnetic field $\Btotal$ so that the direction of the in-plane field component is collinear with the current direction.}
\end{figure*}

In this Letter, we perform magnetotransport experiments on a 2DES confined at the MgZnO/ZnO heterointerface by rotating the sample in the magnetic field in small steps (ca. 0.25$^\circ$) from $\theta=0$ up to $\theta$ approaching 90$^\circ$ and thus carefully exam Landau level crossing events. The sample has a van-der-Pauw geometry with 8 indium Ohmic contacts soldered at the corners and at the side centers of the samples as shown in Fig.~\ref{Fig1}(c). The mobility of the heterostructure is about 300,000~cm$^2$/Vs and its electron density $n$~=~2.05$\times$10$^{11}$cm$^{-2}$ \cite{Falson2011}. The measurements are done in a $^3$He refrigerator at $T\approx~$500mK. The sample is mounted on a single axis rotation stage, that allows varying \textit{in-situ} the angle $\theta$ between the magnetic field and the normal to the 2DES plane. The zero tilt angle is adjusted by rotating the sample in 0.5~T magnetic field until the maximum value of Hall voltage is achieved. For a fixed tilt angle $\theta$, the longitudinal resistance $\Rxx$ and the Hall resistance $\Rxy$ are recorded as a function of a total magnetic field. The measurements are done using a standard lock-in measurement technique at 11.3~Hz and an excitation current of 100~nA. The tilt angle is deduced by adjusting the slope of the Hall voltage as well as by adjusting the pronounced minima of the longitudinal resistance. Changes of charge carrier density after each $^3$He recondensation are not encountered. 

Figure~\ref{Fig1}(a) is the color rendition plot of the longitudinal resistance $\Rxx$ as a function of $B_{\bot}$ and $1/\cos(\theta)$ with red (blue) areas representing high (low) $\Rxx$ values. The  Landau level integer filling factors $\nu$ are noted on top of the panel. Firstly, we affirm that the observed resistance changes are associated with the crossing of spin-resolved Landau levels. We start with drawing in Fig.~1(b) the fan diagram of spin-resolved Landau levels for a fixed $B_{\bot}$ in an increasing $\Btotal$ represented by $1/\cos(\theta)$-axis and assume for the moment the spin susceptibility being constant. For a given $\Bperp$, the energy of spin-up $\uparrow$ (spin-down $\downarrow$) electron states decreases (increases) in an increasing $\Btotal$ as indicated by red (blue) lines in Fig.~1(b). For the discussion below, we introduce the notation ($N$,$\uparrow$/$\downarrow$) to characterize the electron state.  Such a diagram visualizes the possible level crossing events, which show up in the experiment as an increase of $\Rxx$ at the respective integer filling factors. Figure~\ref{Fig1}(b) exemplifies $\Rxx$-traces for $\nu=5$ and 6 as a function of $1/\cos(\theta)$, extracted from Fig.~\ref{Fig1}(a). Horizontal lines indicate the position of the corresponding chemical potentials. When the chemical potential lies within the gap of the energy spectrum, $\Rxx$ tends toward zero. Whenever two levels cross at the chemical potential, the resistance rises. Accordingly the traces exhibit characteristic sharp resistance peaks and thus signal both the crossing of spin-resolved Landau levels and the change of the overall 2DES polarization. For the integer filling factor $\nu=5$, the diagram implies two crossing events; one is the crossing of ($N=1,\downarrow$) with ($N=3,\uparrow$) and the other is the crossing of ($N=0,\downarrow$) with ($N=4,\uparrow$). Experimentally, two resistance peaks are observed and an index $j=2$ and $j=4$ can be assigned to each of them, respectively. Interestingly, the $\Rxx$ trace for $\nu=6$ shows also two coincidence events with $j=3$ and 5, although the Landau level fan diagram suggests that three such events should occur. We note though that the experimental trace shows a tail of the resistance peak, which can be associated with $j=1$ at around $1/\cos(\theta)\approx 1$. Such behavior implies the enhancement of spin susceptibility by so much that the first coincidence has already occurred even at $\theta=0$. This is in-line with previous reports~\cite{TsukazakiPRB, KozukaPRB}. 
This analysis allows also identifying the electronic states at the chemical potential in each region in Fig.~\ref{Fig1}(a) and thus represents the map of electronic states in the magnetic field. 

\begin{figure}[!thb]
\includegraphics{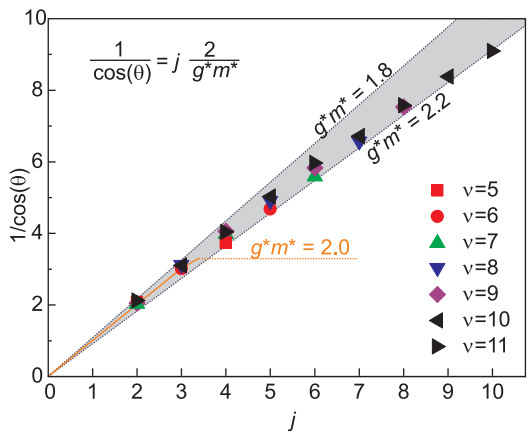}
\caption{\label{Fig2} Estimation of spin-susceptibility for several Landau level integer filling factors. The equation in the inset describes the linear dependence with a slope $g^{\ast}m^{\ast}=$2.0 for $j\leq3$. The gray shaded area indicates a range of $g^{\ast}m^{\ast}$ of 1.8 (upper edge) to 2.2 (lower edge).}
\end{figure}

The level crossing takes place whenever the Zeeman energy is a multiple integer of the cyclotron energy, i.e. $j\cdot E_\text{cyc}=E_\text{z}$; an integer number $j$ characterizes the difference between Landau level orbital momenta $N_{\uparrow}$-$N_{\downarrow}$. Taking into account that $\Bperp=\Btotal\cdot\cos(\theta)$, the coincidence condition can be rewritten as $\frac{g^{\ast}m^{\ast}}{2}=j\cos(\theta)$, which is the expression frequently used to estimate the spin-susceptibility. Accordingly, Fig.~\ref{Fig2} plots $1/\cos(\theta)$, at which the level crossing happens, versus the coincidence index $j$ for several Landau level integer filling factors. For $j\leq3$ all points fall onto a straight line, which passes through the origin and has a slope corresponding to $g^{\ast}m^{\ast}=2.0$, i.e. no significant dependence of $g^{\ast}m^{\ast}$ on the filling factor~\cite{Nicholas1988}. This result is also consistent with our previous measurements, which were done up to $j=3$~\cite{TsukazakiPRB, KozukaPRB}. However, in the larger angles explored in this work, we see that for $j>3$ the points deviate from the single linear dependence and systematically shift to a smaller $1/\cos(\theta)$ for larger $j$'s. This suggests a non-linear process active in the electron system and departure from a level crossing diagram in Fig.~1(b). 

\begin{figure}[!thb]
\includegraphics{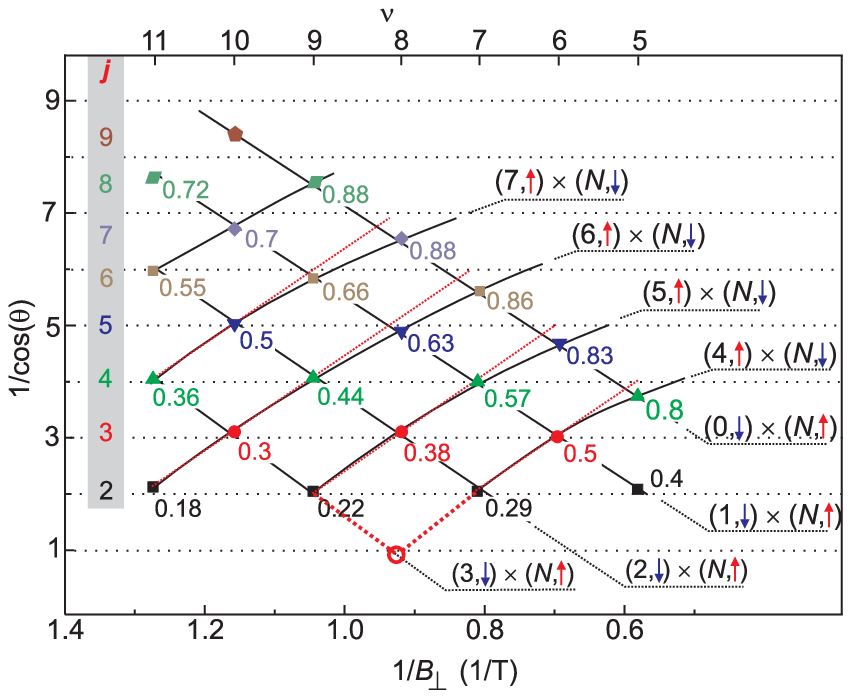}
\caption{\label{Fig3} Position of observed level crossings (closed symbols) at integer filling factors $\nu$ and tilt angles $\theta$ are extracted from Fig.~1(a). The solid lines are guides to the eye, which connect particular coincidence events. A dashed red line emphasizes a systematic shift of coincidence events to smaller $1/\cos(\theta)$. Open circle at the bottom is one of the  $j=1$ coincidence points obtained by extrapolating experimental data points. It lies at $1/\cos(\theta)<1$ and thus emphasizes that the first Level crossing event has occurred in a purely perpendicular magnetic field. The value of net spin polarization $P$ is given at each crossing point.}
\end{figure}

To study the level crossing in more detail, Fig.~\ref{Fig3} summarizes the level crossing events that can be unambiguously identified from gap closing at integer filling factors. Both color and shape of symbols group the coincidence events with the same $j$. The solid curves are guides to the eye, which connect particular crossing events. The lines, which go from top left to bottom right, represent the crossing of Landau levels ($N, \downarrow$) with fixed $N$ with the electronic states ($N^{\prime},\uparrow$). The lines, which run from bottom left to top right, indicate the events when ($N, \uparrow$) electronic levels with $N$~=~4, 5, 6 etc. cross ($N^{\prime},\downarrow$) electronic states with an arbitrary $N^{\prime}$. Firstly, such a map clearly shows that the coincidence events with $j=1$ are absent; a red open circle at $1/\cos(\theta)<1$ denotes the position of hypothetical coincidence event $j=1$ by extrapolating the experimental results. More importantly, we find that the crossing events with the same $j$-index have a tendency to take place at a smaller $\theta$, i.e. a smaller $1/\cos(\theta)$, for smaller $\nu$'s. It appears to correlate well with the electron spin polarization, whose value, depicted in Fig.~\ref{Fig3} next to each crossing point, can be estimated as $P=(n_{\uparrow}-n_{\downarrow})/(n_{\uparrow}+n_{\downarrow})=j/\nu$. Hence, the larger is $P$, the smaller is $\theta$ at which the crossing takes place. Effectively, this corresponds to the increase of $\chi$ with an increasing net spin polarization.

\begin{figure}[!thb]
\includegraphics{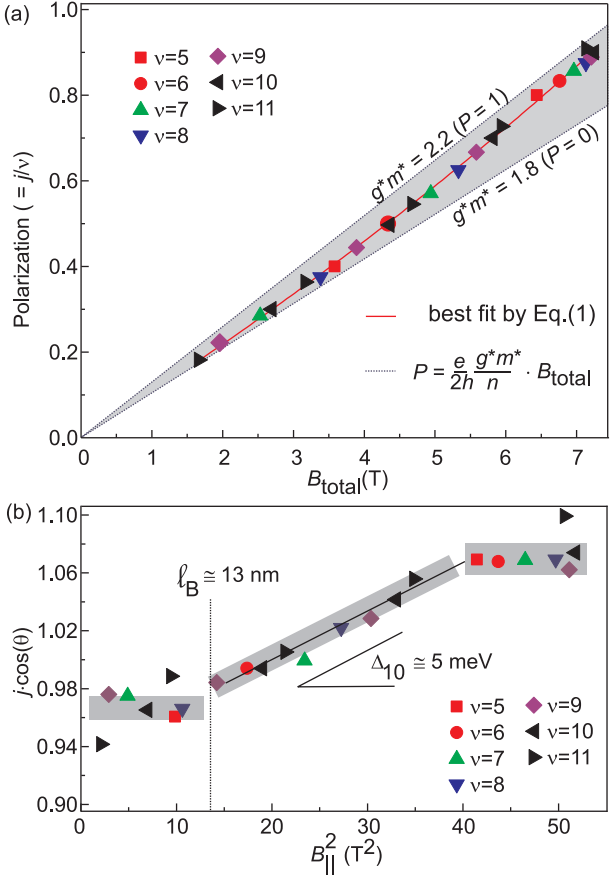}
\caption{\label{Fig4} (a) Spin polarization vs. total magnetic field. The polarization of the electron system is estimated from Fig.~2 as $j/\nu$, where $j$ is the coincidence index. The non-linearity indicates a polarization dependent spin-susceptibility. The experimental points are bound by (dashed) lines of constant susceptibility corresponding to $P=0$ and $P=1$, respectively. (b) Analysis of in-plane field effect on crossing of spin-resolved Landau levels. It provides the lower bound for subband separation $\Delta_{10}=$5~meV and an estimate for the extension of subband electron wave function 13~nm. }
\end{figure}

We scrutinize the role of the spin polarization $P$ by plotting it as a function of $\Btotal$ in Fig.~\ref{Fig4}(a). This plot covers a wide spin polarization range, which otherwise is achieved only in InSb-based 2DES~\cite{Hirayama}, and shows a non-linear dependence of $P$ on $\Btotal$.  
This non-linearity is captured by using Eq.~(1) and assuming a phenomenological relation $\chi=\chi_0+\Delta\chi\cdot P$, where $\chi_0$ is the spin-susceptibility at zero field and $\Delta\chi$ describes the rate of $\chi$ enhancement with $P$~\cite{Hirayama, Nedniyom2009, Zhu2003}. Then in total we obtain:
\begin{equation}
P=\frac{\chi_0 \Btotal}{n-\Delta\chi \Btotal}.
\end{equation}

This dependency describes well the $P-\Btotal$ non-linearity in Fig.~\ref{Fig4}(a) as represented by the best fit (red line) with parameters $\chi_0=1.77\pm0.02$ and $\Delta\chi=0.38\pm0.03$ both in $e/2h$-units. Accordingly, the experimental points fall into the area bound by two lines representing a constant 2DES susceptibility with $\chi_0$, i.e. $P=0$, and with $\chi_0+\Delta\chi$, i.e. $P=1$. The corresponding lines are also shown in Fig.~\ref{Fig2}, where they also represent the bounds for the experimental data and demonstrate the error one does when using coincidence technique to evaluate $\chi$. 

\begin{table}[!h]
\caption{\label{Table1} Comparison of spin susceptibility(in units $e/2h$) for ZnO, AlAs and GaAs based 2DESs with charge carrier density corresponding to the Wigner-Seitz radius $r_s=8$. The data for GaAs and AlAs are extracted from Ref.~\cite{Zhu2003} and \cite{Gokmen2007}, respectively. Here, $\chi_\text{band}$ is the band susceptibility and $P$ is the net spin polarization.}
\begin{ruledtabular}
\begin{tabular}{ccccc}
  2DES &$\chi_\text{band}$&$\chi=\chi_0+\Delta\chi\cdot P$&$\Delta\chi/\chi_0$&$\chi/\chi_\text{band}$\\ \hline
 ZnO & 0.6 & $1.77+0.38\cdot P$ & 21$\%$ &$2.95+0.63\cdot P$  \\
 AlAs & 0.92 & $2.76+0.36\cdot P$ &13$\%$ &$3+0.4\cdot P$ \\
 GaAs & 0.029 & $0.10+0.075\cdot P$ &75$\%$& $3.6+2.57\cdot P$ \\
 \end{tabular}
\end{ruledtabular}
\end{table}

Table~1 compares the spin-susceptibilites in ZnO, AlAs and GaAs-based high mobility 2DESs with the same Wigner-Seitz interaction parameter $r_s\approx 8$. The estimated value $\chi_0=$1.77 in ZnO is a 3-fold enhancement compared to the band value $\chi_\text{band}=0.6$ in ZnO and thus is comparable with $\chi_0$-enhancement in AlAs. While the relative change of spin susceptibility $\Delta\chi/\chi_0$ is larger in ZnO than in AlAs, GaAs shows the largest relative change. It suggests that the coincidence method for the $\chi$-estimation in very diluted GaAs-based 2DESs are more strongly affected than in ZnO or AlAs.

We discuss now whether the $\chi$ enhancement in ZnO is the result of the in-plane magnetic field coupling to the electron orbital motion. 
Such a coupling is known to mix the sub-bands of the confinement potential and to enhance the electron effective mass, which in turn contributes to $\chi$ enhancement~\cite{ParallelFieldTheory, TutucExp, Kozlov2011}. When only the coupling between the ground and the first excited state of the quantized subbands in the confinement potential is considered, the mass is enhanced by a factor $1+\frac{1}{2}\Large(\frac{\hbar\omega_{\|}}{\Delta_{10}}\Large)^2$, where $\omega_{\|}=e\Bparallel/m^{\ast}$ with $\Bparallel$ being the in-plane magnetic field component and $\Delta_{10}$ is the splitting between the subbands. Then the coincidence condition attains an in-plane field dependence $j\cdot\cos(\theta)=\frac{1}{2}g^{\ast}m^{\ast} \approx 1+\frac{1}{2}\Large(\frac{\hbar e/m^{\ast}}{\Delta_{10}}\Bparallel\Large)^2$. This model does not include the electron correlation effects and thus provides the upper bound of the in-plane field effect. Figure~\ref{Fig4}(b) analyzes this model and plots $j\cdot \cos(\theta)$ versus $\Bparallel^2$.  
For small $\Bparallel$, no noticeable field dependence is observed, indicating no significant effect of the orbital coupling. However, starting at $\Bparallel^2\approx 15~T^2$ a linear dependence persists up to $\Bparallel^2$~=~40~T$^2$. In this regime the effect of the in-plane field can account at most for 8$\%$ of mass enhancement. Thereafter no field dependence is observed again, suggesting a saturation of the in-plane field coupling effect and no further electron mass enhancement.  Therefore, $\chi$-dependence discussed above originates primarily from electron correlation effects, with some (8$\%$) contribution from the electron mass enhancement caused by in-plane magnetic field coupling. 

For the sake of analysis, we exploit the constructed model of in-plane field effect and estimate several parameters of the electron confinement potential, which were not probed so far through the transport experiment in MgZnO/ZnO-heterostructures. Firstly, the slope of the linear dependence yields the lower bound for the separation between the two sub-bands of the confinement potential $\Delta_{10}\cong$ 5~meV. It is larger than the Fermi energy of this 2DES 1.6~meV and confirms the single band population. Secondly, the coupling of $\Bparallel$ to the orbital motion is effective if the radius of electron orbital motion, characterized by the magnetic length $\lB$, becomes comparable with the extension of the sub-band wavefunction. Therefore, we estimate the extension of the wavefunction to be 13~nm from the value of the field at which the linear dependence sets on. 

In conclusion, the electrical transport experiments on MgZnO/ZnO heterostructure in a tilted magnetic field mapped out the sequence of electronic states in this novel heterostructure. We measured for the first time in ZnO the polarization dependent spin susceptibility, namely $\chi=1.77+0.38\cdot$$P$ for $r_s \approx 8$, and thus reveal the deviation from the Pauli susceptibility. This observation along with the renormalization of electron spin susceptibility at zero field supports 2DES in ZnO being an interacting Fermi liquid. Our result demonstrates that the position of Landau level crossings depends on the net spin polarization in a strongly interacting electron system and hence highlights the  limitations of the simple coincidence method for estimating the spin susceptibility. The measurement of the net spin polarization dependence on the total magnetic field is an alternative way to estimate the spin-susceptibility.

We would like to thank J. H. Smet, D. Zhang, B. Friess and L. Tiemann for fruitful discussions. 
This work was partly supported by Grant-in-Aids for Scientific Research (S) No. 24226002 from MEXT, Japan, "Funding Program for World-Leading Innovative R$\&$D on Science and Technology (FIRST)" Program from the Japan Society for the Promotion of Science (JSPS) initiated by the Council for Science and Technology Policy. 

\end{document}